\documentclass[onecollarge,runningheads]{FPhys}       % onecolumn "king-size"

\smartqed  % flush right qed marks, e.g. at end of proof

\usepackage{amssymb,color}
\usepackage[normalem]{ulem}
\usepackage{cite}

\usepackage{graphicx}

\newcommand{\bra}[1]{\langle{#1}|}
\newcommand{\ket}[1]{|{#1}\rangle}

\newcommand{\beq}{\begin{equation}}
\newcommand{\eeq}{\end{equation}}
\newcommand{\beqa}{\begin{eqnarray}}
\newcommand{\eeqan}{\end{eqnarray*}}
\newcommand{\beqan}{\begin{eqnarray*}}
\newcommand{\eeqa}{\end{eqnarray}}

\newcommand{\eq}[1]{Eq.~(\ref{#1})}

\newcommand{\sfc}[2]{\mbox{$\frac{#1}{#2}$}}

\journalname{\small{Found Phys (2015)}}
\DOI{\small DOI: 10.1007/s10701-015-9896-3 \quad(\textcolor{blue}{http://dx.doi.org/10.1007/s10701-015-9896-3}\ )}

\begin{document}

\title{T violation and the Unidirectionality of Time: further details of the interference
}

\titlerunning{Found Phys (2015)}        % if too long for running head

\author{Joan A. Vaccaro % \and Second Author
}


\institute{Joan A. Vaccaro \at
           Centre for Quantum Dynamics, Griffith
           University, 170 Kessels Road, Brisbane 4111, Australia.\\
           \email{J.A.Vaccaro@griffith.edu.au}
}
\date{Received: 31 July 2014 / Accepted: 20 March 2015}

\maketitle

\begin{abstract}
T violation has previously been shown to induce destructive interference between different paths that the universe can take through time and leads to a new quantum equation of motion called bievolution.  Here we examine further details of the interference and clarify the conditions needed for the bievolution equation.

\keywords{CP violation \and T violation \and kaons \and arrow of time \and quantum interference \and quantum foundations} \PACS{03.65.-w \and 11.30.Er \and 14.40.-n}

\end{abstract}

\section{Introduction}

Parity inversion (P), charge conjugation (C) and time reversal (T) are fundamental operations in quantum physics. The discoveries that Nature is not invariant under P, CP and T stand as pillars of 20th century physics that promise far reaching consequences. The implications of the violation of CP invariance for quantum decoherence, Bell inequalities, direction of time, particle decay, uncertainty relations and entropy generation has been explored over the last few decades \cite{Andrianov,Hiesmayr,Berger,Durt,Domenico,Sehgal,Bramon,Bernabeu,Escribano,Go,Gerber,Genovese,Bertlmann,Durstberger,Samal,Garbarino,Barnett,Gisin,Taron,Bertlmann1,Grimus,Bohm,Foadia,Nowakowski,Nowakowski2,Bohm2,Corbett,Squires,Datta,Redhead,Finkelstein,Siegwart,Lindblad,Hall,Raychaudhuri}.  Less attention has been given to the violation of T invariance, which is implied by the violation of CP invariance under the CPT theorem and is also observed directly \cite{Angelopoulos,Lees}.  Nevertheless, this fundamental time asymmetry has been linked to the direction of time \cite{Aiello,Fassarella}.  In addition, I have recently shown that the violation of T invariance in neutral meson systems has the potential to affect the nature of time on a large scale \cite{FPhys} far beyond that previously imagined \cite{Price}.

The large scale effects of the violation of T invariance (hereafter referred to as simply T violation) are seen to arise when modelling the universe as a closed system with no predefined direction of time \cite{FPhys}.  T violation implies that there are actually two versions of the Hamiltonian, $\hat H_F$ and $\hat H_B$, one associated with evolution in each direction of time. To avoid prejudicing one direction of time over the other, the time evolution of the system is modelled as a quantum walk comprising a superposition of all possible paths that zigzag forwards and backwards in time. The size of each step in time is fixed at the Planck time. Interference due to T violation is found to eliminate most of the possible paths leaving a superposition of just two paths: one comprises continuous evolution under $\hat H_F$ in the forwards directions of time and the other comprises continuous evolution under $\hat H_B$ in the backwards directions of time.  This result has been called the {\it bievolution} due to the evolution under two Hamiltonians.  Standard quantum mechanics is recovered in each path.  However subtle features of the interference that were not previously appreciated have come to light.  The purpose of the present paper is to reveal those subtle features and explore their repercussions.  We shall see that the regime under which bievolution is valid is more restricted than previously thought.  However, we shall also find that by reducing the size of the time step appropriately, the newly-found restrictions are avoided.

We begin with a brief review in Sect. 2 and then draw out the subtle details of the interference in Sect. 3.  In Sect. 4 we show how they impose a limitation on the range of validity of the bievolution equation and then we examine a resolution of the problem in Sect. 5.  We end with a discussion in Sect. 6.

\section{Review of earlier work}

We briefly review the results presented in Ref.~\cite{FPhys} where full details can be found.  Our analysis applies to a model universe that is composed of matter and fields on a non-relativistic background of space and time.  This model is imagined to be comparable in scale with the visible portion of the actual physical universe.  One could subdivide the model universe into parts which act as spatial and temporal references for the remainder.  We assume, however, that the model universe itself is not a subdivision of something larger, and so there is no physical system external to it.  Thus, there is no external object that could be used as a reference for an origin or direction of space for the universe and, likewise, there is no external clock or memory device that could be used as a reference for time.  In particular, this means we have no basis for favouring one direction of time evolution of the universe over the other. However, maintaining a description that is unbiased with respect to the direction can become cumbersome and so, for convenience, we arbitrarily label the time evolution in the directions of the positive and negative time axes as ``forward'' and ``backward'', respectively.

Let the evolution of the universe in the forward direction over the time interval $\tau$ be given as $\ket{\psi_{F}(\tau)}=\hat U_{F}(\tau)\ket{\psi_{0}}$ where
\begin{equation}
     \label{eq:defn UF}
     \hat U_{F}(\tau)=\exp(-i\tau \hat H_{F})\ .
\end{equation}
Here and elsewhere we set $\hbar=1$ for convenience. Similarly, let the evolution in the backward direction by the same time interval be  $\ket{\psi_{B}(\tau)}=\hat U_{B}(\tau)\ket{\psi_{0}}$ where
\begin{equation}
     \label{eq:defn UB}
     \hat U_{B}(\tau)=\exp(i\tau \hat H_{B})
\end{equation} 
with $\hat H_{B}=\hat T\hat H_{F}\hat T^{-1}$, $\hat U_{B}(\tau)=\hat T\hat U_{F}(\tau)\hat T^{-1}$ and $\hat T i \hat T^{-1}=-i$ where $\hat T$ is Wigner's time reversal operator \cite{Wigner}.  In other words, $\hat H_F$ and $\hat H_B$ are the generators of time translations in the positive$-t$ and negative$-t$ directions, respectively.

If observers within the universe were to consistently find experimental evidence of one version of the Hamiltonian, say $\hat H_{F}$, it would imply that the time evolution is in the corresponding direction, i.e. in the positive--$t$ direction.  But as our analysis needs to be unbiased with respect to the direction of time, it must include both versions of the Hamiltonian on an equal footing.  
An analogous situation occurs in the double slit experiment: if there is no information regarding which slit the particle goes through, we must add the probability amplitudes for the particle going through each slit.
With this in mind, consider the expressions  $\bra{\phi}U_{F}(\tau)\ket{\psi_{0}}$ and $\bra{\phi
}U_{B}(\tau)\ket{\psi_{0}}$ which represent the probability amplitudes for the universe to evolve from the state $\ket{\psi_{0}}$  to the state  $\ket{\phi}$ via two different paths, each of which corresponds to a different direction of time and associated version of the Hamiltonian.  As we have no reason to favour one direction (or Hamiltonian) over the other, we follow Feynman's sum over paths method and take the total probability amplitude to evolve from $\ket{\psi_{0}}$  to  $\ket{\phi}$   as the sum
\begin{equation}
     \label{eq:sum of amps}
   \bra{\phi}U_{F}(\tau)\ket{\psi_{0}}+\bra{\phi
    }U_{B}(\tau)\ket{\psi_{0}} = \bra{\phi}[U_{F}(\tau)+U_{B}(\tau)]\ket{\psi_{0}}\ .
\end{equation}
As this result is true for all states $\ket{\phi}$, it represents the universe evolving from $\ket{\psi_{0}}$ to the state
\beq
     \label{eq:1 step biev = (U+U)^N}
     \ket{\Psi (\tau)} \equiv \left[\hat U_{F} (\tau)+\hat U_{B} (\tau)  \right]\ket{\psi _{0}}\ .
\eeq
We call the process described by \eq{eq:1 step biev = (U+U)^N}  \textit{symmetric time evolution} over one step in time.
Symmetric evolution in both directions of time has also been explored by Carroll, Barbour and co-workers \cite{Chen,Carroll,Barbour} in different contexts. Here it is simply a result of avoiding any bias in the direction of time. 

The extension to the case where there are many different possible paths is straightforward: {\it the total probability amplitude for the universe to evolve from one given state to another is proportional to the sum of the probability amplitudes for all possible paths through time between the two states.} We must be careful, however, to weigh all paths in the sum uniformly.  We can do this in a recursive manner by replacing $\ket{\psi_0}$ in \eq{eq:sum of amps} with $\ket{\Psi (\tau)}$ to give the sum of the amplitudes for evolving from  $\ket{\Psi (\tau)}$ to  $\ket{\phi}$ via two different paths in time.  The symmetric time evolution of $\ket{\Psi(\tau)}$ that results is given by
\beq
    \label{eq:2 step biev = (U+U)^N}
    \ket{\Psi (2\tau)} \equiv \left[\hat U_{F} (\tau)+\hat U_{B} (\tau)  \right]\ket{\Psi(\tau)}
\eeq
which, according to \eq{eq:1 step biev = (U+U)^N}, is the two-step symmetric evolution of $\ket{\psi_0}$:
\beq
    \ket{\Psi (2\tau)}=\left[\hat U_{F} (\tau)+\hat U_{B} (\tau)  \right]^2\ket{\psi _{0}}\ .
\eeq
Repeating this argument $N$ times results in the $N$-step symmetric evolution of $\ket{\psi_0}$ as
\begin{equation}
     \label{eq:N step biev = (U+U)^N}
     \ket{\Psi (N\tau)}=\left[\hat U_{F} (\tau)+\hat U_{B} (\tau)  \right]^{N}\ket{\psi _{0}}\ .
\end{equation}
This was shown to be equal to
\begin{equation}
     \label{eq:N step biev = S_N-n,n}
     \ket{\Psi (N\tau)}=\sum\limits_{n=0}^N {\hat S_{N-n,n}} \;\ket{\psi_{0}}
\end{equation}
where $\hat S_{N-n,n}$ is the sum of $(^N_n)$ terms each of which is a unique ordering of $N-n$ factors of $\hat U_{B}(\tau)$ and $n$ factors of $\hat U_{F}(\tau)$.

Any given ordering of factors, such as $\cdots\hat U_{F}(\tau)\hat U_{F}(\tau)\hat U_{B}(\tau)$, represents the evolution along a path that zigzags in time, going one step backward, two forward, etc. and so the expression $\bra{\varphi}\cdots$ $\hat U_{F}(\tau)\hat U_{F}(\tau)\hat U_{B}(\tau)\ket{\psi_{0}}$ represents the corresponding probability amplitude for evolving from $\ket{\psi_{0}}$ to $\ket{\varphi}$ over that path.  This means that the expression $\bra{\varphi}\hat S_{N-n,n}\ket{\psi_{0}}$ represents the total probability amplitude for evolving from $\ket{\psi_{0}}$ to $\ket{\varphi}$ along $(^N_n)$ different paths in time. The evolution along different paths can destructively interfere leading to a small, or even zero, total probability amplitude.  A clock device that is part of the Universe and constructed of matter that obeys T invariance, would evolve in time by the same net amount of $(N-2n)\tau$ for each path represented by $\bra{\varphi}\hat S_{N-n,n}\ket{\psi_{0}}$.  The use of T-invariant matter in the construction of the clock device avoids any ambiguity in the length of time intervals for different directions of time.

Reordering $\hat S_{m,n}$ so that all the $\hat U_{B} (\tau)$ factor are to the left of the $\hat U_{F} (\tau)$ factors gives
\beqa
    \label{S_m,n=U_BU_F sum commutator}
    \hat S_{m,n} = \hat U_{B} (m\tau) \hat U_{F} (n\tau)\sum\limits_{v=0}^m
    \cdots \sum\limits_{\ell =0}^s \sum\limits_{k=0}^\ell
    \exp \left[    {(v+\cdots +\ell +k)\tau ^{2}[\hat {H}_{F} ,\hat {H}_{B} ]} \right]
\eeqa
to order $\tau^3$.  There are $n$ summations on the right side and $[\hat A,\hat B]$ is the commutator of $\hat A$ and $\hat B$.  It is useful to express the summations in \eq{S_m,n=U_BU_F sum commutator} in terms of the eigenvalues of the Hermitian operator  $i[\hat {H}_{F} ,\hat {H}_{B} ]$.  Writing the operator $i[\hat {H}_{F} ,\hat {H}_{B} ]$ in terms of its eigenbasis as
\beq \label{eq:commutator with rho}
  i[\hat {H}_{F} ,\hat {H}_{B} ] = \int \lambda \rho(\lambda)\hat \Pi(\lambda) d\lambda\ ,
\eeq
where $\hat\Pi(\lambda)$ is an operator with unit trace such that $\rho(\lambda)\hat\Pi(\lambda)$ is a projection operator that projects onto the manifold of eigenstates of $i[H_F,H_B]$ with eigenvalue $\lambda$, and $\rho(\lambda)$ is a density function representing the degeneracy of eigenvalue $\lambda$ with $\int_\Lambda \rho(\lambda)d\lambda$ being the number of eigenstates in the interval $\Lambda$,  then yields \cite{FPhys}
\beq   \label{eq:S}
    \hat S_{m,n} = \hat U_{B}(m\tau)\hat U_{F}(n\tau)\int I_{m,n} (\tau^{2}\lambda)\rho(\lambda)\hat\Pi(\lambda) d\lambda
\eeq
where\footnote{Note that $I_{m,n}(\tau^{2}\lambda)$ here is equivalent to the function $I_{m,n}(\lambda)$ in Ref.~\cite{FPhys}.}
\beq \label{eq:interference function defined}
  I_{m,n}({z})= \sum\limits_{v=0}^m {\cdots \sum\limits_{\ell =0}^s
       {\sum\limits_{k=0}^\ell {\exp \left[ {-i\,(v+\cdots +\ell +k){z}} \right]}}}
\eeq
is an {\it interference function} which has been shown to be symmetric with respect to the indices $n$ and $m$, i.e. $I_{m,n}({z})=I_{n,m}({z})$. After some algebraic manipulation \cite{FPhys} we find
\beqa \label{eq:interference function as product}
     I_{N-n,n}({z})
          &=& \exp[-in(N-n){z}/2]{\prod_{q=1}^n
          {\frac{\sin [(N+1-q){z}/2]}
          {\sin (q{z}/2)}}} \ .
\eeqa

In Ref.~\cite{FPhys} the value of $\tau$ was fixed at the Planck time $\tau\approx 5\times 10^{-44}$s and the degeneracy $\rho(\lambda)$ was taken to be Gaussian distributed about a mean of $\lambda=0$.  It was shown that $I_{N-n,n}(\tau^{2}\lambda)$ was sharply peaked at $\lambda=0$ for $n\gg 0$ and  $n\ll N$ from which it was argued that the integral in \eq{eq:S} becomes approximately proportional to the projection operator $\rho(0)\hat\Pi(0)$, and so if $\rho(0)\hat\Pi(0)\ket{\psi_0}=0$ then the corresponding term $\hat S_{N-n,n}$ is missing from the sum in \eq{eq:N step biev = S_N-n,n}.  An absence of the term indicates destructive interference between the associated paths through time.  It was concluded that for suitable states of the Universe, destructive interference eliminates all paths except for those that are either continuously backward (corresponding to $n\approx 0$) or continuously forward (corresponding to $n\approx N$).  The result is the {\em bievolution equation}
\beq
   \label{eq:bievolution eq}
    \ket{\Psi(N\tau)}\approx\left[\hat U_{F}(N\tau)+\hat U_{B}(N\tau)\right]\ket{\psi _{0}}
\eeq
for sufficiently large values of $N$.  However, subtle details of the interference function $I_{N-n,n}(\tau^{2}\lambda)$ that were not previously appreciated place restrictions on the range of validity of this result.  In the remainder of this paper we explore those details and elucidate their repercussions.

\section{Details of the interference function}

We noted previously that $|I_{N-n,n}({z})|$ has a central maximum at ${z}=0$ for which \cite{FPhys}
\beq  \label{eq:central maximum of I}
     I_{N-n,n}(0) = \left({}^N_n\right)\ .
\eeq
We shall refer to this as the {\it principle maximum} of $|I_{N-n,n}({z})|$ and any other maxima as {\it subsidiary maxima}.  We can find other features of the interference function $I_{N-n,n}({z})$ by determining the points along the ${z}$ axis where it takes on particular values.  To do this, consider the factors in the iterated product of \eq{eq:interference function as product} each of which comprises a numerator of the form $f_{\rm num}(N,q,{z})=\sin[(N+1-q){z}/2]$ and a denominator of the form $f_{\rm den}(q,{z})=\sin(q{z}/2)$.  We now deduce three properties as follows.

(i) The {\it zeroes} of $I_{N-n,n}({z})$ occur at the points away from the origin where one of the numerators is zero, that is, for $f_{\rm num}(N,q,{z})=0$ and $z\ne 0$ which is satisfied by
\beq  \label{eq:zeroes of I}
   {z}=\frac{2m\pi}{N+1-q}
\eeq
for $q=1,2,\ldots,n$ and non-zero integer $m$.

(ii) The interference function has a {\it modulus of unity} at the points where each factor in the iterated product has a modulus of unity.  This occurs at a nonzero value of ${z}$ where $f_{\rm num}(N,q,{z})=\pm f_{\rm den}(q,{z})$ for all values of $q$ in the range $1,2,\ldots,n$.  Writing this condition as $\sin[(N+1){z}/2-q{z}/2]=\pm\sin(q{z}/2)$ shows that a set of solutions is given by $(N+1){z}/2=m\pi$, or
\beq  \label{eq:points where I is unity (A)}
   {z}=\frac{2m\pi}{N+1}
\eeq
for non-zero integer $m$.  The iterated product in \eq{eq:interference function as product} can be rearranged to give a product in which the numerators are multiplied in reverse order as follows
\[
    \prod_{q=1}^n \frac{f_{\rm num}(N,q,{z})}{f_{\rm den}(q,{z})}=\prod_{q=1}^n \frac{f_{\rm num}(N,n+1-q,{z})}{f_{\rm den}(q,{z})}\ .
\]
Each factor on the right side has a modulus of unity for $\sin[(N-n){z}/2+q{z}/2]=\pm\sin(q{z}/2)$, which is satisfied by $(N-n){z}/2=m\pi$, i.e. by
\beq  \label{eq:points where I is unity (B)}
   {z}=\frac{2m\pi}{N-n}\ .
\eeq

(iii) The modulus of the interference function becomes {\it maximal} at the points where all the numerators satisfy $f_{\rm num}(N,q,{z})\approx\pm 1$ and all the denominators are small, i.e. where $f_{\rm den}(q,{z})\ll 1$ and so $\sin(q{z}/2)\approx q{z}/2 \ll 1$ to first order in $q{z}/2$.  At these points the modulus of the interference function is given approximately as
\beq  \label{eq:bound on peaks of I}
     |I_{N-n,n}({z})|
          \approx \prod\nolimits_{q=1}^n\frac{1}{q{z}/2}
          =\frac{1}{n!}\left(\frac{2}{{z}}\right)^n\ .
\eeq
This simple expression will give the approximate magnitude of a maximum in $|I_{N-n,n}({z})|$ provided we can specify the points along the ${z}$ axis where the maxima occur.  These points can be found for $n\ll N$ as follows. We restrict our attention to the case where each denominator $f_{\rm den}(q,{z})$ for $q=1,2,\ldots,n$ varies over much less than one cycle with respect to ${z}$, i.e. $0 < z \ll 4\pi/n$.  The condition on the numerators implies that $\sin[(N+1-q){z}/2]\approx\pm 1$ which, given that $n\ll N$ and so $q\ll N$, is satisfied by
\beq  \label{eq:condt for max}
   (N+1){z}/2\approx(2m+1)\pi/2
\eeq
for integer $m$.  We shall take first subsidiary maximum along the positive $z$ axis as the one that occurs after the first zero of $I_{N-n,n}({z})$ which, according to \eq{eq:zeroes of I}, is for $z>2\pi/N$. Equation~(\ref{eq:condt for max}) then gives the position of the $m$-th subsidiary maximum of $|I_{N-n,n}({z})|$ along the positive ${z}$ axis as
\beq  \label{eq:points where I is large}
    {z}_m=\frac{(2m+1)\pi}{(N+1)}
\eeq
for positive integer $m$.  Combining this with the condition $z \ll 4\pi/n$ for the denominators to be small implies that \eq{eq:bound on peaks of I} is valid for the $m$-th maximum provided
\beq   \label{eq:condition on approximation for large I}
   \frac{(2m+1)\pi}{(N+1)} \ll \frac{4\pi}{n}\ .
\eeq

To summarise, this analysis implies that $|I_{N-n,n}({z})|$ ranges from (i) zero at the points given by \eq{eq:zeroes of I} through (ii) unity at the points given by \eq{eq:points where I is unity (A)} and \eq{eq:points where I is unity (B)} and to (iii) subsidiary maxima that are bounded by \eq{eq:bound on peaks of I} at the points \eq{eq:points where I is large}.

\begin{figure}[h]  %%%%%%%%%%%%%%%%%%%%%%%%%%%%%%%%%%%%%
\begin{center}\includegraphics[width=11.0cm]{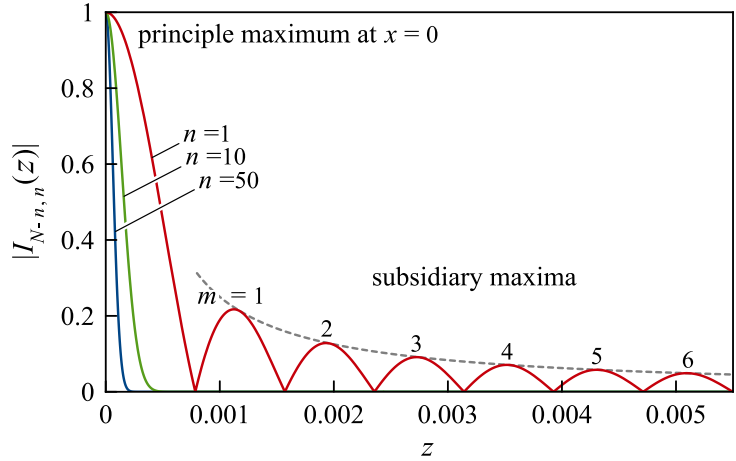}
\caption{(Colour online) The modulus of $I_{N-n,n}({z})$ for $N=8000$ and $n=1$ (red curve), $n=10$ (green curve) and $n=50$ (blue curve).  For clarity, each curve has been scaled by dividing $|I_{N-n,n}({z})|$ by $(^N_n)$ to give a value of unity at ${z}=0$.  The dotted gray line represents the bound on the subsidiary maxima for $n=1$ given by \eq{eq:bound on peaks of I}.
\label{fig:linear I}}
\end{center}
\end{figure}

The principal maximum is significantly larger than the bound on the subsidiary maxima given by \eq{eq:bound on peaks of I} for values of $n$ that are significantly different from $1$ and $N$ where $N\gg 1$.  Figure~\ref{fig:linear I} compares the relative heights of the principal and subsidiary maxima for $N=8000$ and shows that the subsidiary maxima for the cases $n=10$ and $n=50$ are negligible (and below the resolution of the figure) in comparison to the principal maximum.

    %%%%%%%%%%%%%%%%%%%%%%%%%%%%%%%%%%%%%

\begin{figure}[h]  %%%%%%%%%%%%%%%%%%%%%%%%%%%%%%%%%%%%%
\begin{center}\includegraphics[width=11.0cm]{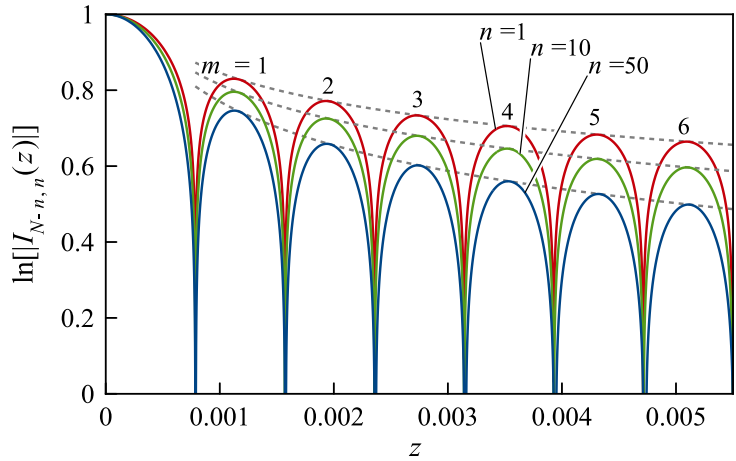}
\caption{(Colour online) Logarithm of $|I_{N-n,n}({z})|$ for $N=8000$ and $n=1$ (red curve), $n=10$ (green curve) and $n=50$ (blue curve).  Again, for clarity, each curve has been scaled by dividing $\ln[|I_{N-n,n}({z})|]$ by $\ln[(^N_n)]$ to give a value of unity at ${z}=0$.  The dotted gray lines represents the bounds given by \eq{eq:bound on peaks of I}.
\label{fig:logarithm I}}
\end{center}
\end{figure}

\begin{figure}[h]  %%%%%%%%%%%%%%%%%%%%%%%%%%%%%%%%%%%%%
\begin{center}\includegraphics[width=11.0cm]{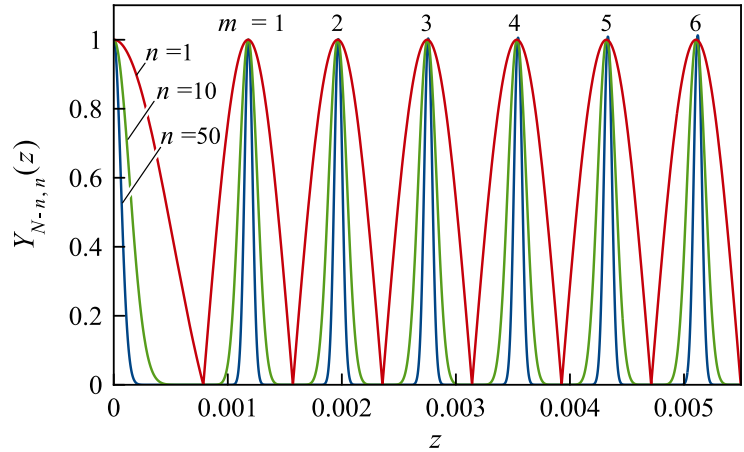}
\caption{(Colour online) The scaled interference function $Y_{N-n,n}({z})$ for $N=8000$ and $n=1$ (red curve), $n=10$ (green curve) and $n=50$ (blue curve).
\label{fig:scaled I}}
\end{center}
\end{figure}

Nevertheless the subsidiary maxima play a significant role that was not appreciated previously.  In Fig.~\ref{fig:logarithm I} we plot the natural logarithm of $|I_{N-n,n}({z})|$ for the same cases as in Fig.~\ref{fig:linear I}.  Also plotted are the bounds (dotted grey curves) given by \eq{eq:bound on peaks of I} which accurately estimate the magnitudes of the subsidiary maxima.  Knowing the magnitude of the principle maximum, from \eq{eq:central maximum of I}, and the bound on the subsidiary maxima, from \eq{eq:bound on peaks of I}, allows us to scale the interference function so that all maxima are of equal height using the following scaling function,
\[
   F_{N,n}({z}) = \left\{\begin{array}{l} ({}^N_n)\ , \mbox{ for } x\le \frac{2\pi}{N+1}\\
                                         \frac{1}{n!}\left(\frac{2}{{z}}\right)^n, \mbox{ otherwise}\ .
                       \end{array}\right.
\]
Figure~\ref{fig:scaled I} shows the scaled interference function $Y_{N-n,n}({z})$ given by
\beq  \label{eq:scaled interference fn Y}
  Y_{N-n,n}({z})=\frac{I_{N-n,n}({z})}{F_{N,n}({z})}
\eeq
for the same cases as in Fig.~\ref{fig:logarithm I}.  Notice that the height of the sixth subsidiary maxima (i.e. for $m=6$) for the case $n=50$ is slightly larger than unity due to the approximate nature of the bound in \eq{eq:bound on peaks of I} used for the scaling function (despite condition \eq{eq:condition on approximation for large I} being satisfied).

    %%%%%%%%%%%%%%%%%%%%%%%%%%%%%%%%%%%%%

    %%%%%%%%%%%%%%%%%%%%%%%%%%%%%%%%%%%%%

\begin{figure}[h]  %%%%%%%%%%%%%%%%%%%%%%%%%%%%%%%%%%%%%
\begin{center}\includegraphics[width=11.0cm]{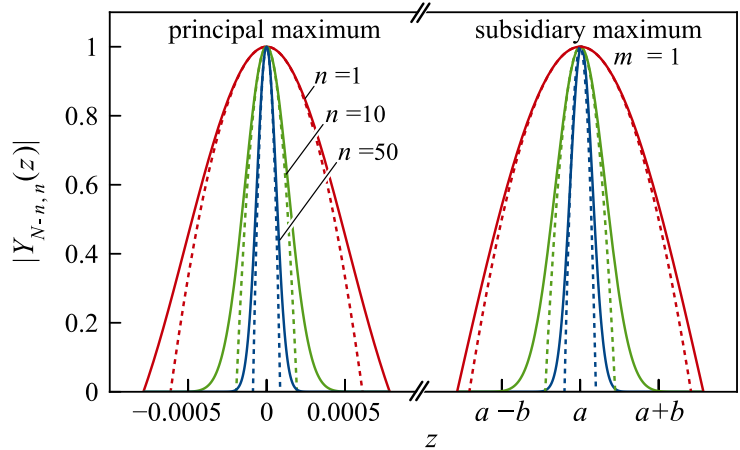}
\caption{(Colour online) Quadratic approximations (dotted curves) given by Eqs.~(\ref{eq:quadratic approx for subsidiary max}) and (\ref{eq:quadratic approx for principle max}) to the scaled interference function $Y_{N-n,n}({z})$ (solid curves) for the principle and first subsidiary maximum.  The plots are for $N=8000$ and $n=1$ (red), $n=10$ (green) and $n=50$ (blue). Note that different scaling is used for the ${z}$ axis for the principal and subsidiary maxima.  The subsidiary maximum is centered on ${z}=a$ where $a=3\pi/(N+1)$ and $b=0.00025$.
\label{fig:quadratic}}
\end{center}
\end{figure}

We can estimate the widths of the subsidiary maxima by fitting a quadratic function as follows. We set
\[
    {z}={z}_m+\epsilon\ ,
\]
where ${z}_m$ is the position of the $m$-th maximum given by \eq{eq:points where I is large}, and express the right side of \eq{eq:interference function as product} as a function of $\epsilon$.  The quadratic approximation will only be valid over a range of values of $\epsilon$ that is much less than half the distance between consecutive maxima, i.e. for
\beq  \label{eq:condition on epsilon}
   |\epsilon| \ll \frac{\pi}{N+1}\ .
\eeq
In the limit of large $N$, we find that the factors of the iterated product in \eq{eq:interference function as product} can be approximated for $q\ll N$ and $qz_m\ll 1$ by
\beq   \label{eq:factor near peak}
   \frac{\sin[(N+1-q)({z}_m+\epsilon)/2]}{\sin(q({z}_m+\epsilon)/2)}
      \approx  \frac{\pm\cos[(N+1-q)\epsilon/2]}
                {\sin(q{z}_m/2)}
\eeq
where the sign depends on the value of $m$ and we have made use of \eq{eq:condition on epsilon} in the denominator. The arguments of the trigonometric functions on the right of \eq{eq:factor near peak} are much smaller than $\pi$, and so expanding them to second order yields
\beq
   \frac{\sin[(N+1-q)({z}_m+\epsilon)/2]}{\sin(q({z}_m+\epsilon)/2)}
      \approx \pm \frac{ 1-(N+1-q)^2\epsilon^2/8}
                {q{z}_m/2}\ .
\eeq
Substituting this result into \eq{eq:interference function as product} gives the modulus of the interference function as
\[
    |I_{N-n,n}({z}_m+\epsilon)|
          \approx \prod_{q=1}^n\frac{ 1-(N+1-q)^2\epsilon^2/8}
                {q{z}_m/2}
\]
which is approximately
\[
    |I_{N-n,n}({z}_m+\epsilon)|
          \approx \frac{ 1-\sum_{q=1}^n(N+1-q)^2\epsilon^2/8}
                {n!({z}_m/2)^n}
\]
to second order in $\epsilon$.  Evaluating the summation and retaining terms of order $(N+1)$ or higher  then gives
\beq  \label{eq:quadratic approx for subsidiary max}
    |I_{N-n,n}({z}_m+\epsilon)|
          \approx \frac{1}{n!}\left(\frac{2}{{z}_m}\right)^n\left[1-\epsilon^2\frac{n(N-n)(N+1)}{8}\right]
                \ .
\eeq
We found using the same technique in Ref.~\cite{FPhys} that the quadratic approximation to the principle maximum was given by
\beq   \label{eq:quadratic approx for principle max}
   |I_{N-n,n}(\epsilon)|\approx \left({}^N_n\right)\left[1-\epsilon^2\frac{n(N-n)(N+1)}{24}\right]\ .
\eeq
Figure~\ref{fig:quadratic} illustrates the closeness of these approximations for the principle and first subsidiary maxima for various values of $n$.

    %%%%%%%%%%%%%%%%%%%%%%%%%%%%%%%%%%%%%

The widths of the peaks can be estimated from the points where the right sides of Eqs.~(\ref{eq:quadratic approx for principle max}) and (\ref{eq:quadratic approx for subsidiary max}) are zero.  The corresponding values of $\epsilon$ are
\beqa
   \epsilon_{\rm prin}&=&\sqrt{\frac{24}{n(N-n)(N+1)}}\\
   \epsilon_{\rm sub}&=&\sqrt{\frac{8}{n(N-n)(N+1)}}
\eeqa
for the principle and subsidiary maxima, respectively.  These values are bounded by
\beqa
   \label{eq:width prin max}
\epsilon_{\rm prin}&\lesssim&\frac{4.9}{\sqrt{k}N}\\
   \label{eq:width sub max}
   \epsilon_{\rm sub}&\lesssim&\frac{2.8}{\sqrt{k}N}
\eeqa
where $k$ is the lessor of $n$ and $N-n$.

\section{Implications for the bievolution equation}

\begin{figure}  %%%%%%%%%%%%%%%%%%%%%%%%%%%%%%%%%%%%%
\begin{center}\includegraphics[width=11.0cm]{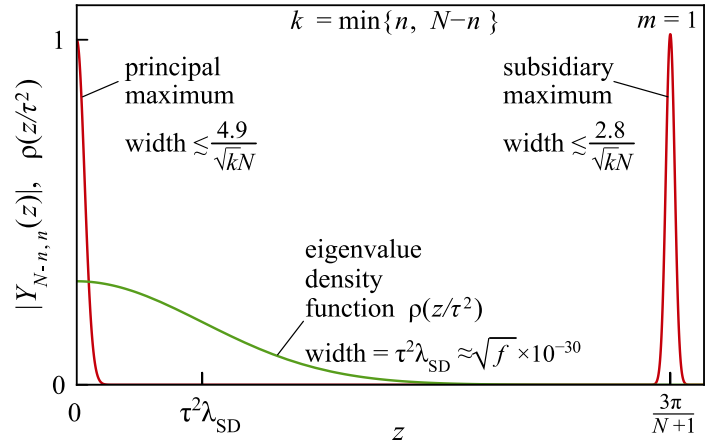}
\caption{(Colour online) Sketch comparing the scaled interference function of $|Y_{N-n,n}({z})|$ (red curve) with the eigenvalue density function $\rho({z}/\tau^2)$ (green curve).
\label{fig:comparison}}
\end{center}
\end{figure}

We have now uncovered sufficient information about the subsidiary maxima to be able to explore their implications for the derivation of the bievolution equation \eq{eq:bievolution eq}.  Combining Eqs.~(\ref{eq:N step biev = S_N-n,n}) and (\ref{eq:S}) gives an expression for the state $\ket{\Psi (N\tau)}$ in terms of the interference function $I_{m,n} (\tau^{2}\lambda)$ and the density function $\rho(\lambda)$ as
\beq  \label{eq:psi with interference and density fns}
\ket{\Psi (N\tau)}=\sum\limits_{n=0}^N  \hat U_{B}[(N-n)\tau)]\hat U_{F}(n\tau)\left[\int I_{N-n,n} (\tau^{2}\lambda)\rho(\lambda)\hat\Pi(\lambda) d\lambda\ \ket{\psi_{0}}\right]\ .
\eeq
It is important to keep in mind in the following that the state $\ket{\Psi (N\tau)}$ is not normalized; in particular, to enable a consistent probabilistic interpretation its magnitude needs to be normalized for each value of $N$.  In order for \eq{eq:psi with interference and density fns} to yield the bievolution equation, the terms for values of $n$ that significantly different from $0$ and $N$ must be negligible in comparison to the remaining terms. The only way for this to occur independently of the general nature of the operators $\hat U_{B}(\cdot)\hat U_{F}(\cdot)$ is if the expression in square brackets vanishes for these terms.  To see how this might happen consider the two functions $I_{N-n,n}(\tau^{2}\lambda)=I_{N-n,n}(z)$ and $\rho(\lambda)=\rho({z}/\tau^2)$ that appear in the integral of \eq{eq:psi with interference and density fns}, where $z=\tau^2\lambda$. Fig.~\ref{fig:comparison} illustrates the degree to which their product contributes to the integral.  The scaled interference function $Y_{N-n,n}(z)$ given by \eq{eq:scaled interference fn Y} is used in the figure to emphasize the role of the subsidiary maxima and, as in Ref.~\cite{FPhys}, we assume the eigenvalue density function $\rho(\lambda)$ is Gaussian and centered on $\lambda=0$ with a standard deviation of $\lambda_{\rm SD}\approx \sqrt{f}\times 10^{57} s^{-2}$ where $f$ represents the fraction of particles in the universe that violate T invariance.  We also assume that the state $\ket{\psi_{0}}$ satisfies the {\it nonzero eigenvalue condition} \cite{FPhys}:
\beq \label{eq:nonzero eigenvalue condt}
  \rho(0)\hat\Pi(0)\ket{\psi_{0}}=0\ ,
\eeq
i.e. it has no projection onto the subspace of eigenstates of $i[H_F,H_B]$ with eigenvalue zero.

    %%%%%%%%%%%%%%%%%%%%%%%%%%%%%%%%%%%%%

First take the case where the position of the $m=1$ subsidiary maximum is significantly greater than the width of the density function, i.e. where
\beq   \label{eq:condition for bievolution eq}
    \frac{3\pi}{N+1} \gg \tau^2\lambda_{\rm SD}\ .
\eeq
This places an upper bound on the value of $N$.  In this case, provided $0\ll n\ll N$ the integral in \eq{eq:psi with interference and density fns} is approximately proportional to the projection operator $\rho(0)\hat\Pi(0)$ which projects onto the subspace with zero eigenvalues.  Due to the nonzero eigenvalue condition \eq{eq:nonzero eigenvalue condt}, the corresponding terms in \eq{eq:psi with interference and density fns} are negligible.  The bievolution equation then follows, as reported in Ref.~\cite{FPhys}.

Next take the case where the value of $N$ violates the upper bound in \eq{eq:condition for bievolution eq} and, to be specific, let
\beq
    \frac{3\pi}{N+1} = \tau^2\lambda_{\rm SD}\ .
\eeq
Again consider the range $0\ll n\ll N$ and, due to the nonzero eigenvalue condition, note that $\rho(0)\hat\Pi(0)\ket{\psi_{0}}=0$. The position of the subsidiary maximum is now centered on $z=\tau^2\lambda_{\rm SD}$ and so the expression in square brackets in \eq{eq:psi with interference and density fns} is approximately proportional to $\rho(\lambda_{\rm SD})\hat\Pi(\lambda_{\rm SD})\ket{\psi_{0}}$, which is nonzero in general. This implies that the corresponding terms in \eq{eq:psi with interference and density fns} are not necessarily zero and so the bievolution equation no longer holds.  Moreover, further consideration along these lines will show that the bievolution equation is not recovered, in general, for larger values of $N$.

Thus, \eq{eq:condition for bievolution eq} represents a condition that must be satisfied in order for the bievolution equation to hold.  Expressing the condition as an upper bound on the total time $N\tau$ gives
\beq   \label{eq:bound on total time}
     N\tau \ll \frac{3\pi}{\tau\lambda_{\rm SD}}\approx \frac{10^{-13}~s}{\sqrt{f}}\ .
\eeq
For the right side to be 10 billion years, the fraction $f$ of particles in the visible universe that are kaon-like and contribute to T violation needs to be $10^{-61}$.  Given it contains an estimated $10^{80}$ proton-like particles, this means that the visible universe needs to contain much less than a mole of kaon-like particles in order for the mechanism to explain the direction of time.

\section{Reducing the value of $\tau$}

\begin{figure}[b]  %%%%%%%%%%%%%%%%%%%%%%%%%%%%%%%%%%%%%
\begin{center}\includegraphics[width=11.0cm]{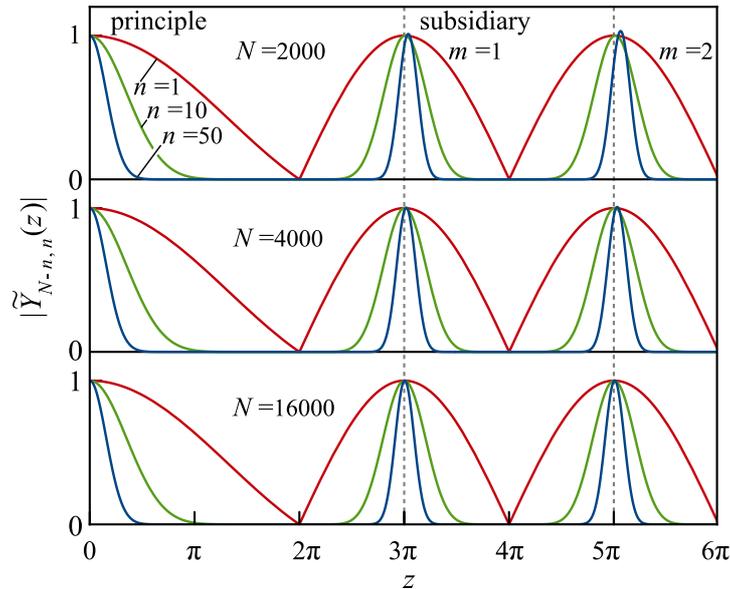}
\caption{(Colour online) Plots of $|\widetilde{Y}_{N-n,n}({z})|=|{Y}_{N-n,n}(\frac{z}{N+1})|$ for different values of $N$ verifying that the position of the subsidiary maxima are relatively fixed at $z=(2m+1)\pi$.
\label{fig:reducing tau}}
\end{center}
\end{figure}    %%%%%%%%%%%%%%%%%%%%%%%%%%%%%%%%%%%%%

The physical unreasonableness of such a small value of $f$ calls for a review of the approach we have employed. Up to now the size of the step in time, $\tau$, has been fixed at the Planck time.  But \eq{eq:bound on total time} can be satisfied for more reasonable values of $f$ if $\tau$ is allowed to have a much smaller value. Moreover, the preceding analysis shows that condition \eq{eq:condition for bievolution eq} inevitably fails at some value of $N$ because the first subsidiary maximum moves towards the origin of the $z$ axis as the values of $N$ increases.  The positions of all the subsidiary maxima would remain relatively fixed, however, if the size of $\tau$ scales as $1/\sqrt{N+1}$.  For example, replacing $\tau$ in \eq{eq:condition for bievolution eq} with $c/\sqrt{N+1}$, where $c$ is a constant, yields the constraint on the position of the first subsidiary maximum as
\[
    \frac{3\pi}{N+1} \gg \frac{c^2}{N+1}\lambda_{\rm SD}\ ,
\]
which is satisfied, independently of the value of $N$, provided $c^2\lambda_{\rm SD}\ll 3\pi$.  Correspondingly, the argument of $Y_{N-n,n}({z})$ in \eq{eq:scaled interference fn Y} should be replaced with $\tau^2\lambda=c^2\lambda/(N+1)=z/(N+1)$.  This yields a new function of $z$ as follows:
\beq  \label{eq:scaled interference fn Y prime}
  \widetilde{Y}_{N-n,n}({z})=Y_{N-n,n}(\sfc{z}{N+1})\ .
\eeq
Figure~\ref{fig:reducing tau} verifies that the position of the first subsidiary maximum of $|\widetilde{Y}_{N-n,n}({z})|$ converges to
\[
    z=3\pi
\]
for large values of $N$.  As $z=c^2\lambda$, this corresponds to $\lambda=3\pi/c^2$.  Thus, if $\widetilde{Y}_{N-n,n}({z})$ is plotted as a function of $\lambda$, the first subsidiary maximum will be fixed at $\lambda=3\pi/c^2$.  This position can be made far beyond the width $\lambda_{\rm SD}$ of the eigenvalue density function $\rho(\lambda)$  by choosing the value of $c$ accordingly.
That being the case, the integral in \eq{eq:psi with interference and density fns} will be approximately proportional to the projection operator $\rho(0)\hat\Pi(0)$ for $0\ll n\ll N$ and, if the nonzero eigenvalue condition \eq{eq:nonzero eigenvalue condt} holds, the corresponding terms in \eq{eq:psi with interference and density fns} will be negligible.  Hence, the conditions for the bievolution equation \eq{eq:bievolution eq} can be satisfied as $N$ increases indefinitely.  However, the full analysis of the consequences of reducing the value of $\tau$ in this way is beyond the scope of this work and will be explored elsewhere \cite{latest}.

\section{Discussion}

In Ref.~\cite{FPhys} it was argued that destructive interference due to T violation reduces \eq{eq:N step biev = S_N-n,n} to the approximate bievolution equation \eq{eq:bievolution eq}.  The highest degree of accuracy entailed all terms in \eq{eq:N step biev = S_N-n,n} vanishing except for $n=0$ and $n=N$ which imposes the condition on the total time
\beq \label{eq:FPhys condt - stringent}
 N\tau\gg \frac{10^{-13}}{\sqrt{f}}~\mbox{s}\ ,
\eeq
whereas the less accurate form of the bievolution equation
\beq   \label{eq:bievolution eqn - approx}
        \ket{\Psi (N\tau)}=\left[\sum\limits_{n\approx 0} {\hat S_{N-n,n}} + \sum\limits_{n\approx N} {\hat S_{N-n,n}}\right]\ket{\psi_{0}}
\eeq
is satisfied by the less-stringent condition
\beq \label{eq:FPhys condt - less stringent}
 N\tau > 10^{-17}~\mbox{s}\ .
\eeq
However, subtle details of the interference were overlooked.  The present work has revealed the interference function $I_{N-n,n}({z})$ contains subsidiary maxima that were not previously considered.  Their presence was shown to lead to a new condition \eq{eq:bound on total time} that must be met for the bievolution equation to be valid.  The fact that \eq{eq:bound on total time} conflicts with \eq{eq:FPhys condt - stringent} means that the destructive interference cannot be so strong as to eliminate all terms in \eq{eq:N step biev = S_N-n,n} except for $n=0$ and $n=N$, as previously thought.  Instead, the best one can achieve is the more approximate from given by \eq{eq:bievolution eqn - approx}.  In that case the combination of \eq{eq:FPhys condt - less stringent} and \eq{eq:bound on total time} gives the range of values of the total time $N\tau$ over which the bievolution equation is valid as
\[
   10^{-17}~\mbox{s} < N\tau \ll \frac{10^{-13}}{\sqrt{f}}~\mbox{s}\ .
\]
We found that the range of allowed $N\tau$ values is unlikely to extend to the current age of our universe as that would require an unreasonably small proportion of T violating particles.

Nevertheless, we also found that the new condition condition \eq{eq:bound on total time} could be avoided if $\tau$, rather than being fixed at the Planck time, reduces in proportion to $1/\sqrt{N+1}$ as the number of steps $N$ increases. In that case the upper limit to the total time \eq{eq:bound on total time} no longer applies.  However, further analysis of this new approach is beyond the scope of the present work and will be presented elsewhere \cite{latest}.

In conclusion, although we have shown that the effect of destructive interference for fixed $\tau$ has been overstated in previous work \cite{FPhys}, nevertheless, we have also found that the destructive interference can be recovered by modifying the method and allowing $\tau$ to reduce as the number of steps increases.

\end{document}